%% file: manuscript.tex
\documentclass[final,11pt, letterpaper]{article}
\usepackage[noblocks]{authblk}

\newcommand{\params}{\rho IEA}

\makeatletter
\renewcommand\AB@authnote[1]{}
\renewcommand\AB@affilnote[1]{}
\makeatother

\usepackage{commath}
\usepackage[sf,bf,small]{titlesec}
\usepackage{amsmath, amssymb}
\usepackage{array}
\usepackage{multicol}
\usepackage{booktabs}
\usepackage{xcolor}
\usepackage{colortbl}
\usepackage[inline]{enumitem}
\usepackage[pdftex,final]{graphicx}
\usepackage[hidelinks]{hyperref}
\usepackage[labelfont=bf,
            small,
            labelsep=period,
            figurename=Figure]
            {caption}

\usepackage{setspace}
\usepackage[]{lineno}
\usepackage[letterpaper,margin=1.0in]{geometry}
\usepackage[authoryear,sort&compress,comma,square]{natbib}
\usepackage[symbol]{footmisc}
\usepackage{subcaption}

\graphicspath{{}}

\def\fig{figure}
\def\tab{table}
\def\eqn{equation}

\date{}

\title{Non-invasive load measurement in the human tibia via spectral analysis of flexural waves}

\author[ ]{Ali Yawar\thanks{Address correspondence to: aliyawar92@gmail.com.}\thanks{Current affiliation: Amazon Robotics, North Reading, MA, USA. The work in this paper is unrelated to this author's affiliation with Amazon.}} 
\author[ ]{Daniel H. Aslan}
\author[ ]{Daniel E. Lieberman}

\affil[ ]{Department of Human Evolutionary Biology, Harvard University, Cambridge MA, USA}

\begin{document}

\maketitle

\abstract{
Forces transmitted by bones are routinely studied in human biomechanics, but it is challenging to measure them non-invasively, especially outside of laboratory settings.
We introduce a technique for non-invasive, \emph{in vivo} measurement of tibial compressive force using flexural waves propagating in the tibia. Modelling the tibia as an axially compressed Euler-Bernoulli beam, we show that tibial flexural waves have load-dependent frequency spectra. Specifically, under physiological conditions, peak locations in the wave acceleration spectra vary linearly with the compressive force on the tibia and may be used as proxies for the compressive force. We test the validity of this technique using a proof-of-concept wearable system that generates flexural waves via a skin-mounted mechanical transducer and measures the spectra of these waves using a skin-mounted accelerometer. In agreement with beam theory, data from 9 participants demonstrate linear relationships between tibial compressive force and spectral peak location, with Pearson correlation coefficients $r=0.82 - 0.99$ (mean $r=0.93$) for medial-lateral swaying and $r=0.81 - 0.98$ (mean $r=0.93$) for walking trials. This flexural wave-based technique could give rise to a new class of wearable sensors for non-invasive physiological bone load monitoring and measurement, impacting research in human locomotion and sports medicine.}

\section{Introduction}
The human tibia is subjected to a dynamic combination of bending, twisting, and axial loads during locomotion.
In running, the tibia can sustain compressive forces reaching up to 14 times the bodyweight \citep{scott1990internal}.
A majority of the tibial compressive force is attributed to tension from triceps surae muscle contractions during propulsion \citep{matijevich2019ground}.
Tibial compressive loading, like loading in other bones, impacts maintenance \citep{kohrt2004physical}, growth \citep{robling2009mechanical}, and injury \citep{moen2009medial,warden2014management}, and is therefore a key variable to measure in experimental human biomechanics.

Currently, in laboratory or clinical settings, a standard technique to estimate tibial loading during locomotion integrates optical motion capture and instrumented treadmills with inverse dynamics models \citep{winter2009biomechanics}. 
A more invasive method uses strain gauges embedded directly within the bone, with gauge placement limited to the medial diaphysis of the tibia in humans \citep{rubin2001vivo,milgrom2004comparison,yang2011we}.
For in-field use, the primary tools are wearable inertial measurement units (IMUs) or insole pressure sensors which measure bone load proxies like tibial acceleration or ground reaction force \citep{elstub2022tibial, willy2025average}.
But despite decades of research and use, wearable IMU-based proxies are not reliable surrogates for bone loading \citep{willy2025average,xiang2024rethinking,matijevich2019ground}.
Recent machine learning-based approaches may yield better accuracy, but the inherent participant-specific modelling assumptions and lack of interpretability hamper their broad applicability \citep{matijevich2020combining}.  

Here we introduce a non-invasive technique that uses propagating flexural waves in the tibia to measure compressive loading. 
In a beam-like structure like the tibia, flexural wave propagation is governed by material and geometric properties and the axial load borne by the structure \citep{doyle2020wave}.
Tibial flexural waves have previously been used for non-invasive monitoring of material properties \citep{fah1988phase,stussi1988assessment,vogl2016reliability}, but their relationship with tibial load has not been studied.
Modelling the tibia as an axially compressed Euler-Bernoulli beam \citep{doyle2020wave}, we show that its mechanical impedance is load-dependent and consequently flexural waves in the tibia have frequency spectra that change with the axial compressive force on the tibia.
Thus, the force on the tibia may be inferred by tracking variations in the flexural wave spectra.
We demonstrate this technique using a proof-of-concept, non-invasive wearable system that generates propagating flexural waves in the tibia using a skin-mounted audio transducer (or ``tapper'') and measures their spectra using a skin-mounted accelerometer.
We test this system in $N=9$ participants performing medial-lateral swaying (at approximately 0.2~Hz) and walking (at 0.4 m/s) on an instrumented treadmill, and compare our measurements against tibial compressive force measured using inverse dynamics.
The non-invasive and portable nature of our technique has implications for the development of new wearable sensors for in-field measurements and continuous monitoring of bone loading in athletic, clinical, or industrial environments.

\section{Results}
\subsection{Flexural wave propagation model}

\begin{figure}[h!]
    \centering
    \includegraphics[width=0.5\textwidth]{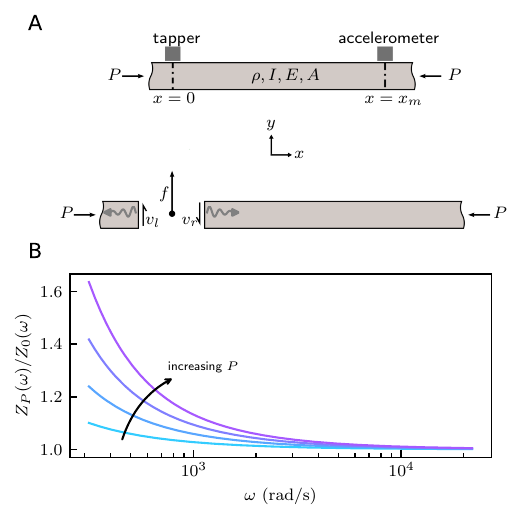}
    \caption{\textbf{Axially compressed Euler-Bernoulli beam model.} \textbf{A}, The tibia is modeled as an infinitely long Euler-Bernoulli beam with density $\rho$, cross-sectional moment of inertia $I$, Young's modulus $E$, and cross-sectional area $A$, supporting a uniform axial force $P$ which is positive in compression. A transducer (or ``tapper'') at $x=0$ applies a transverse force $f(t)$ to the beam, supported by shear forces $v_l, v_r$ on either side of a virtual cut at the origin. As a result, a flexural disturbance propagates symmetrically away from the origin (shown by wavy arrows) which is measured by an accelerometer located at $x_m \gg 0$. \textbf{B}, Plot of the far-field transfer impedance magnitude of the axially loaded beam normalized by the impedance of the zero-load beam, $Z_P(\omega)/Z_0(\omega)$, versus frequency $\omega$. Frequency is on a logarithmic scale. Colours denote a range of axial compressive loads ($0 < P \leq 4000$~N). As load increases, the normalized impedance increases more for lower frequencies than for higher frequencies.
    Parameter values for this plot are based on published values for the human tibia: $\rho=1850$~kg~m$^{-3}$, $E=20$~GPa, $I$ = $10^{-8}~$m$^4$, A = $2.5 \times 10^{-4}$~m$^2$ \citep{rho1996ultrasonic,minns1975geometrical}.}
    \label{fig:theory}
\end{figure}

We model the tibia as a uniform, infinitely long Euler-Bernoulli beam with density $\rho$, cross-sectional moment of inertia $I$, Young's modulus $E$, and cross-sectional area $A$ (figure~\ref{fig:theory}A).
Here, infinite length is equivalent to ignoring wave reflections from the boundaries, so that the resulting solution provides the baseline flexural response of the tibia without any resonance (SI section~\ref{sec:model validity}). The beam supports a uniform axial compressive force $P$. The governing equation for transverse displacement $y(x,t)$ is
\begin{equation}
    EI\frac{\partial^4 y}{\partial x^4} + P\frac{\partial^2 y}{\partial x^2} + \rho A\frac{\partial^2 y}{\partial t^2} = 0,
    \label{eq:EB equation} 
\end{equation}
where $x,t$ are the axial spatial coordinate and time, respectively.
We solve equation~\eqref{eq:EB equation} in the frequency ($\omega$) domain for the right half of the beam ($x\geq 0$) following \citet{doyle2020wave}. At $x=0$, the tapper (figure~\ref{fig:theory}A) applies a transverse force $f(t)$ with Fourier transform $\mathcal{F}(\omega)$. 
By symmetry, this force is supported equally by shear forces on either side of the origin, resulting in the boundary condition $f/2 + v_r =0$ where $v_r = -EI\partial^3y/\partial x^3$ is the shear force on the right half of the beam (details in SI section~\ref{sec:boundary conditions}).
Combining this shear force boundary condition with the condition for symmetric wave generation at the origin $\partial y/\partial x = 0$, equation~\eqref{eq:EB equation} yields the spectrum of transverse displacement at a far-field location $x\gg0$,
\begin{equation}
    \mathcal{Y}(P, \omega, x) = \dfrac{-i\mathcal{F}}{2EI(\xi^3 + \xi\kappa^2)}e^{-i\xi x},
    \label{eq:EB solution}
\end{equation}
where $\xi, \kappa \in \mathbb{R}^+$ are load- and frequency-dependent wavenumbers defined by,
\begin{equation}
    \xi(P,\omega) = \dfrac{\sqrt{P + \sqrt{P^2 + 4\params\omega^2}}}{\sqrt{2EI}},
    \quad \kappa(P,\omega) = \dfrac{\sqrt{-P + \sqrt{P^2 + 4\params\omega^2}}}{\sqrt{2EI}}.
    \label{eq:wavenumbers}
\end{equation}
The transfer impedance of the beam between the origin and a fixed measurement location $x_m \gg 0$ is defined as
\begin{equation*}
    \mathcal{Z}(P,\omega) = \dfrac{\mathcal{F}}{i\omega\mathcal{Y}(P,\omega,x_m)},
\end{equation*}
and its magnitude is found using equation \eqref{eq:EB solution} as
\begin{equation}
    Z_P(\omega) \equiv |\mathcal{Z}(P,\omega)| =  \dfrac{2EI(\xi^3 + \xi\kappa^2)}{\omega}.
    \label{eq:impedance}
\end{equation}
The magnitude of the impedance $Z_P$ normalized by the impedance of the zero-load beam $Z_0$ is plotted against frequency $\omega$ in figure~\ref{fig:theory}B. 
For high frequencies, axial compression has little effect on impedance as $Z_P/Z_0 \approx 1$.
At lower frequencies the compressed beam has a higher impedance than the beam with zero load ($Z_P/Z_0 > 1$).
Thus, lower frequencies are selectively suppressed.

A measurable effect of the beam's load-dependent impedance is that the acceleration (or velocity or displacement) spectrum of propagating flexural waves will shift towards higher frequencies with increasing compressive force.
Consequently, if the magnitude of the acceleration spectrum for the zero-load beam has a peak at the frequency $\omega^*$, impedance change due to a small axial load ($0 < P\ll EA$) will cause a shift in the spectral peak location given by 
\begin{equation}
    \Delta \omega^* = \dfrac{P}{\displaystyle 4\sqrt{\params}} \dfrac{\gamma(\omega^*)}{\omega^{*2}},
    \label{eq:main result linear}
\end{equation}
where $\gamma(\omega^*)$ is a parameter that depends on the sharpness of the spectral peak at zero load (SI section~\ref{sec:peak shift derivation}).
Thus, the spectral peak location shifts linearly with the compressive force on the tibia.
Assuming constant material and geometric properties of the tibia, equation~\eqref{eq:main result linear} may be used to infer relative changes in the tibial compressive force by measuring shifts in peak locations in the flexural wave acceleration spectra.

\subsection{Implementation of the sensor system}

\begin{figure}[h!]
    \centering
    \includegraphics[width=0.7\textwidth]{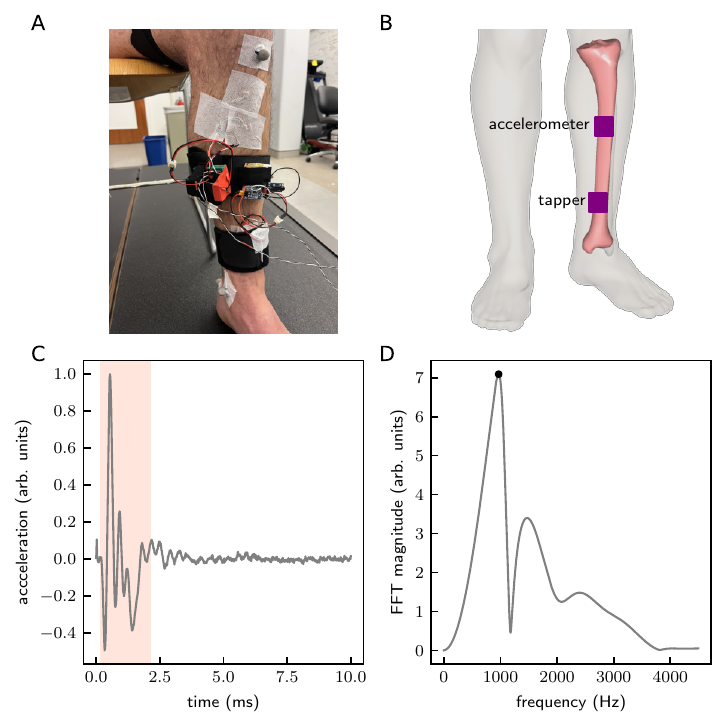}
    \caption{\textbf{Overview of the sensor setup and measurements.} \textbf{A}, Photograph of a participant's left leg showing the system setup and \textbf{B}, approximate locations of the audio transducer (``tapper'') and accelerometer on a schematic of the tibia. The tapper delivers a sequence of impulsive taps at 50~Hz, and the accelerometer measures the resulting response of the tibia. \textbf{C}, The measured transverse acceleration in response to a single tap initiated at $t=0$~s shows a highly damped signal. The signal is sampled at 100~kHz, mean-centered, and normalized by the maximum acceleration. The 2~ms portion of the acceleration signal used for analysis is highlighted in pink. \textbf{D}, Frequency domain representation of the highlighted acceleration signal from panel C via the magnitude of the Fast Fourier Transform (FFT), showing the peak of the magnitude spectrum (solid black dot). Subsequent analyses use the locations of spectral peaks to infer the compressive force on the tibia. The leg and tibia schematics are from BodyParts3D \citep{mitsuhashi2009bodyparts3d}, used under a CC-BY-SA license.}
    \label{fig:intro}
\end{figure}

The above derivation considers the beam's flexural response to a transverse force input $f(t)$.
If the duration of $f(t)$ and the resulting response of the tibia are short (akin to a ``tap''), the tibial compressive force may be assumed constant over the duration of the response.
If a periodic sequence of identical taps is applied to the tibia, the flexural wave generated in response to each tap will correspond to the tibial compressive force at the instant of the tap, assuming that successive waves can be temporally separated.
By measuring the response to successive taps, tibial loading may be continually tracked at a sampling rate equal to the rate of tapping.
In our setup, a skin-mounted tapper near the medial malleolus delivers a sequence of impulsive taps at 50~Hz to generate a train of flexural waves in the tibia.
These waves are measured using a skin-mounted accelerometer at the mid-tibia (figure~\ref{fig:intro}A,B).
The flexural wave acceleration response to a single tap shows that the highly damped response settles well before the subsequent tap is initiated (figure~\ref{fig:intro}C).
Peak locations in the magnitude of the acceleration spectra are then extracted and used to infer the compressive force on the tibia (figure~\ref{fig:intro}D). 

\subsection{In vivo measurement of flexural wave spectra}

\begin{figure}[h!]
    \centering
    \includegraphics[width=0.5\textwidth]{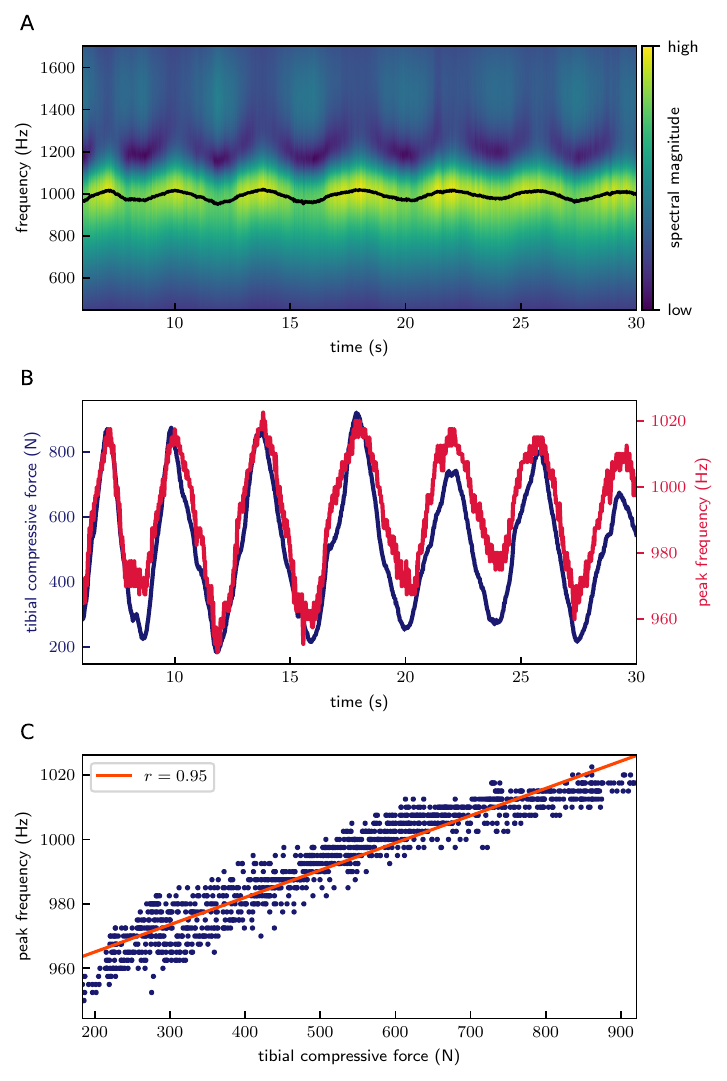}
    \caption{\textbf{Tibial compressive force measurement in medial-lateral swaying.} \textbf{A}, Spectrogram of the flexural wave acceleration measured in response to a periodic sequence of taps delivered at 50~Hz. Each vertical slice is the spectrum of the acceleration response to a single tap with the magnitude of the spectrum denoted by colour. Peak locations in the spectrogram form a ridge (black). Data shown here span 6 to 30~s of the trial, as the first 6~s are used for calibration and not used in subsequent analyses. \textbf{B}, The reference tibial compressive force measured using inverse dynamics (blue) shows strong correspondence with the peak frequency ridge extracted from the spectrogram (red) over the trial duration. \textbf{C}, Scatter plot of the two measurements over the entire trial shows a linear dependence with a Pearson correlation coefficient $r=0.95$. The linear regression line is shown in red.}
    \label{fig:results-static}
\end{figure}

Using our wearable system, we measure flexural wave spectra in the left tibia of participants as they perform medial-lateral swaying on a stationary instrumented treadmill, during which the tibia undergoes an approximately periodic compressive loading ($\sim$ 0.2~Hz).
The time-varying spectra of acceleration from one participant are shown as a spectrogram in figure~\ref{fig:results-static}A.
Each vertical slice shows the magnitude of the acceleration spectrum in response to a single tap.
Peak locations in the spectrogram form a ridge fluctuating around 1000~Hz (black curve).
The spectrogram spans 6 to 30~s of the trial, as the first 6~s are used for calibration to locate the region of interest within which the peak ridge is extracted.
Calibration data are not used in subsequent analyses.
The extracted peak ridge corresponds strongly with the reference tibial compressive force measured using inverse dynamics (figure~\ref{fig:results-static}B), with a Pearson correlation coefficient, $r=0.95$ (figure~\ref{fig:results-static}C).

\begin{figure}[h!]
    \centering
    \includegraphics[width=0.5\textwidth]{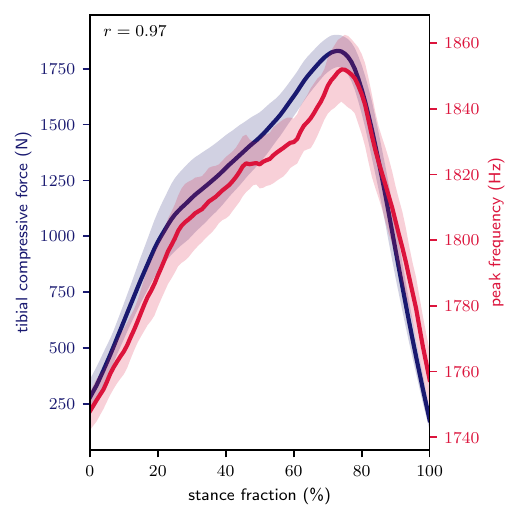}
    \caption{\textbf{Tibial compressive force measurement during walking.} Ensemble averaged data from the walking trial of one participant, showing the correspondence between the reference tibial compressive force measured using inverse dynamics (blue) and the peak frequency ridge extracted from flexural wave spectra (red). The mean correlation coefficient between the two measurements across all stance phases is $r=0.97$. Force and frequency data are averaged over all stance phases and normalized over the period of stance with heel strike at 0\% and toe off at 100\%. Solid curves show the means and the shaded regions span one standard deviation above and below the mean.}
    \label{fig:results-walking}
\end{figure}

We also measure the flexural wave spectra in participants walking at 0.4~m/s. 
Figure~\ref{fig:results-walking} shows data from one participant, comprising the ensemble mean and standard deviation of the extracted spectral peak ridge and the reference tibial compressive force, both segmented and normalized to the duration of stance.
The mean correlation coefficient across all stance phases for this participant is $r=0.97$. 

As summarized in table~\ref{tab:subject-wise-results}, correlation coefficients for the swaying trials range from 0.82 to 0.99 (mean $r=0.93$) across all participants.
In the walking trials, correlation coefficients range from 0.81 to 0.98 (mean $r=0.93$).
The corresponding data figures for all participants are provided in the SI figures~\ref{fig:swaying-data-supplement} and \ref{fig:walking-data-supplement}.

\begin{table}[t]
\centering
\begin{tabular}{c c c} 
\hline
Participant & Swaying $r$ & Walking $r$\\ 
\hline
1 & 0.95 & 0.97 \\ 
2 & 0.99 & 0.98 \\ 
3 & 0.88 & 0.91 \\ 
4 & 0.96 & 0.81 \\ 
5 & 0.92 & 0.94 \\ 
6 & 0.82 & 0.93 \\ 
7 & 0.83 & 0.90 \\ 
8 & 0.94 & 0.94 \\ 
9 & 0.95 & 0.82 \\ 
\hline
mean $r$ & $0.93$ & $0.93$ \\ 
\hline
\end{tabular}
\caption{Pearson correlation coefficients across all participants for the swaying and walking trials.}
\label{tab:subject-wise-results}
\end{table}

\section{Discussion}

Our results demonstrate that propagating flexural waves in the tibia may be used to measure relative changes in the tibial compressive force.
With increasing compressive force, the impedance of the tibia changes such that lower flexural wave frequencies are suppressed more than higher frequencies (figure~\ref{fig:theory}B).
Consequently, the wave spectrum shifts to higher frequencies with increasing compression.
Using a skin-mounted tapper near the distal end of the tibia, we generate a train of propagating flexural waves whose spectra are measured using a skin-mounted accelerometer at the mid-tibia.
In agreement with predictions of the beam model, we see a strong linear relationship between compressive force on the tibia and peak locations in the flexural wave spectra (figures~\ref{fig:results-static},\ref{fig:results-walking}, SI figures \ref{fig:swaying-data-supplement}, \ref{fig:walking-data-supplement}).
While it is not the primary focus of this study, we note that the phase of the flexural wave acceleration spectrum is also load-dependent (equation~\eqref{eq:EB solution}).
At a specific frequency and at a fixed measurement location, the phase changes linearly with the compressive force on the tibia (SI equation~\eqref{suppeq: phase}).
However, in our implementation, phase measurements showed poorer linear correlations with load for most trials, making the phase spectrum a less suitable target as compared to the magnitude spectrum.

As a reference measurement of tibial compressive force, we use the sum of ground reaction force projected along the tibia and calf muscle force, similar to previous studies \citep{matijevich2019ground,holowka2021human,stearne2016foot}. Calf muscle force (assumed equal to the Achilles tendon tension) is defined as the sagittal plane ankle moment divided by the Achilles tendon moment arm.
While this simplified model captures essential characteristics of the true tibial compressive force \citep{matijevich2019ground}, it ignores contributions of the other large muscles spanning the tibia like the tibialis posterior/anterior, and the toe flexors/extensors, all of which contribute to the compression of the tibia during dynamic movement.
Detailed musculoskeletal models and direct comparisons against bone-embedded sensors in future studies will help evaluate more precise limits on the absolute accuracy of our technique.

Our choices of system parameters such as the locations of the tapper and accelerometer, tapper force profile, sampling rate, etc. were selected to accomplish a simple proof-of-concept implementation of the technique.
We placed the tapper near the ankle as it was the most comfortable site for participants in pilot trials and placed the accelerometer away from the tapper and the knee to satisfy far-field requirements of our model.
While the flexural response of the tibia has been found to not vary significantly with distance from the wave source \citep{vogl2016reliability}, future implementations may still benefit from empirical studies to determine combinations of these system parameters for optimal signal to noise ratio.
Furthermore, since our participant cohort was limited to a narrow range of ages and comprised only healthy participants, larger scale studies are necessary before the validity of our proposed technique can be generalized to the population at large.
Additional experimentation at higher speeds and under high impact movements is needed to further ascertain this generalizability.
We expect the measurements to be robust to disturbance from vibration even in high-impact running, where the noise is typically at frequencies lower than our frequencies of interest ($\sim 30$~Hz, \citep{chadefaux20193d,horvais2019cushioning}).
  
There are a few limitations to this technique.
First, skin mounted accelerometer measurements are not feasible on body locations that have significant subcutaneous fat, muscles, or tendons.
Cadaveric studies show that skin-mounted acceleration measurements are dampened in comparison with bone-mounted measurements \citep{nokes1984vibration}.
While preloading the accelerometers with springs may reduce the amount of damping by skin and other soft tissues \citep{nokes1984vibration}, the associated signal distortion may render the measurement unusable, especially if consistent preload cannot be maintained. 
Second, the spectrum of flexural waves in our model only depends on the \emph{net} compressive force on the tibia and cannot resolve stress gradients across the tibial cross-section, which may arise due to bending or torsion of the tibia during load bearing \citep{yang2014torsion}.
Thus, our technique only measures one aspect of how the tibia is loaded.  
Finally, to prevent interference, the separation between successive taps must be sufficient to allow the response of the tibia from one tap to become quiescent before the next tap is delivered.
This settling time depends on the characteristics of the bone and the apparatus and sets a fundamental upper limit on the rate at which tibial compressive forces can be sampled with this technique.
In our setup, the response of the tibia became quiescent about 6~ms after a tap for most participants, suggesting that taps may be delivered at rates of up to about 160~Hz. 

The structure of the acceleration spectra showed wide variability across participants (SI figures \ref{fig:swaying-data-supplement}).
The likely causes of this variability are differences in bone material properties and geometry, and inter-participant inconsistency in the interfaces between the tibia and the tapper or accelerometer.
This in turn affected which frequency ranges in the spectrum were most sensitive to load. 
So, we performed calibration using a 6~s subset of the data from each trial to identify a region of interest (ROI) in the spectrogram containing peak locations that best correlate with tibial load.
In the remainder of the trial, peaks were extracted form this calibrated ROI.
While the results are promising, the necessity of a calibration step using laboratory equipment is not practical for a wearable device meant for field use. 
Successful use of the technique in a portable setup will require the development of a simpler calibration method, or a modification such that calibration is no longer necessary.
Such considerations are common in a broad range of non-invasive sensors that measure proxies for physiological loading such as tendon shear wave speed \citep{martin2018gauging}, muscle thickness \citep{jin2024estimation}, and surface EMG \citep{disselhorst2009surface}.

Our technique may serve as a target for commercial wearable sensor development.
In this study the tapper and accelerometer were tethered to data acquisition equipment, but the requisite processing may be accomplished by hardware worn on the body.
We highlight two engineering challenges for a practical implementation of this technique in a wearable sensor. First, is the design of soft goods that ensure consistent skin interfaces during prolonged movement, both for the accelerometer and for the tapper. 
Second, the small form factor necessary for portability would require choosing sampling rates and tapping force parameters such that real-time processing and power demands may be met with lightweight batteries and microcontrollers.  
We anticipate several potential use cases for a wearable sensor that measures tibial compression.
Such a sensor could enable in-field monitoring of tibial loading, adding to the repertoire of wearables-based measurements of injury proxies \citep{baker2025estimating,baker2024predicting,xu2025data}.
Directly measuring tibial loading could also enable prospective studies on athletic populations in natural settings, which may help answer persistent questions on the causes and prevention of repetitive stress injuries \citep{reshef2012medial}.
Recently, wearable IMU-based systems have been used to improve movement form in athletes by providing live feedback during physical activity \citep{van2024effect}. 
The near real-time tibial load measurements enabled by our technique could be used similarly and may help mitigate the accuracy limitations of IMU-based systems \citep{rokhmanova2024imu}.
In wearable lower-limb exoskeletons, estimates of tibial compressive force may serve as an additional signal with which high-level gait phase estimators can disambiguate user intent \citep{baud2021review}.

\section{Conclusion}
In this work, we theoretically derive how flexural waves propagating in the tibia respond to the axial compressive force on the tibia.
Drawing on this relationship, we introduce a simple, non-invasive wearable system that generates propagating flexural waves in the tibia and measures their spectra to infer the compressive force.
Our proof-of-concept system demonstrates the feasibility of using this technique in human participants and opens the door for its future development into a portable wearable sensor.

\section{Methods and materials}
\subsection{Participants}

We recruited $N=9$ participants (5 male, 4 female; age $31.3\pm5.6$~years, body mass $71.0\pm14.1$~kg, mean$\pm$s.d.) with no recent history of lower limb injury. The study protocol was approved by the Committee on the Use of Human Subjects at Harvard University. Each participant provided written informed consent. 

\subsection{Data collection}

\subsubsection*{Kinematic and kinetic data}
Retro-reflective markers were placed on 9 bony landmarks on the left limb: greater trochanter, medial and lateral tibial condyles, tibial tuberosity, medial and lateral malleoli, posterior calcaneus inferior to the insertion location of the Achilles tendon, and heads of the first and fifth metatarsals. Kinematic data were collected using an eight-camera motion capture system (Qualisys AB, Gothenburg, Sweden) at 200~Hz. Force data were collected using a split-belt instrumented treadmill (Bertec, Columbus, OH, USA) at 2000~Hz. Both kinematic and force data were filtered using a zero-phase, fourth-order Butterworth lowpass filter with a cutoff frequency of 25~Hz. 
Rigid body segments were defined for the shank (tibia) using markers on the tibial condyles and malleoli, and for the foot using markers on the first and fifth metatarsal, and posterior calcaneus.
The long axis of the shank was defined as the line joining the virtual ankle joint center (mid-point of the malleoli) to the virtual knee joint center (mid-point of the tibial condyles). 
The foot and shank segments were connected at the virtual ankle joint center, where the ankle moment was computed.
Inverse dynamics calculations were performed in Visual3D (C-Motion, Inc., Moyds, MD, USA). For the medial-lateral swaying trial, the treadmill belt was stopped and participants were asked to stand in a neutral posture with one foot on each belt and then slowly shift their weight from one foot to another without lifting either foot from the treadmill. Participants swayed at approximately 0.2~Hz. 
For the walking trials, participants walked at 0.4~m/s. The slow walking speed ensured that the tethering cables were not a tripping hazard and did not tug on the fragile electronic devices worn by the participants.

Following \citet{matijevich2019ground}, the reference compressive force on the tibia was estimated as the sum of two forces: i) the ground reaction force projected onto the long axis of the shank, and ii) calf muscle force. 
Calf muscle force was defined as the sagittal plane ankle moment (in the shank frame of reference) divided by the moment arm of the Achilles tendon.
Rather than assuming a constant Achilles tendon moment arm as in \citet{matijevich2019ground}, we used the antero-posterior distance between the ankle joint center and the posterior calcaneus marker, similar to \citet{holowka2021human}.
Time instants when the vertical ground reaction force was below 20~N or when the center of pressure was posterior to the ankle joint center were excluded from the analyses. 

\subsubsection*{Audio transducer (``tapper'')}
We used a 1~W bone conduction audio transducer or ``tapper'' (Product \# 1674, Adafruit Inc., NY, USA) to deliver periodic force perturbations (``taps''). 
A square wave (10~V, 50~Hz) generated in LabChart v8.1.30 (ADInstruments) and amplified using a 2.5~W audio amplifier (PAM8302, Adafruit Inc., NY, USA) was used to drive the tapper.
A 3.7~V LiPo battery was used as the power supply for the amplifier. 
Shielded coaxial cables were used to transmit the square wave signal to the amplifier to minimize interference. The tapper was placed near the distal end of the tibia about 4cm proximal to the medial malleolus, resting on the flat antero-medial face of the tibial shaft.
It was attached to skin using adhesive gauze tape (BSN Cover-Roll Stretch, Hamburg, Germany), and then further secured using velcro fabric bands typically used with wearable IMUs.
Flexural waves were generated by taps from both the rising and falling edges of the square wave, but only the rising edge taps were considered for analysis in this study to ensure consistency in the responses, as rising and falling edge movements of our off-the-shelf tapper were not always identical.

\subsubsection*{Accelerometer}
Flexural wave acceleration was measured using a piezoelectric voice accelerometer (VA1200, Vesper Inc., MA, USA). 
The surface mount accelerometer was soldered to a small, custom designed flexible printed circuit board (6.4~mm by 6.6~mm ``Flex'' style PCB, OSH Park LLC, OR, USA).
A 3V alkaline battery pack was used as the power supply.
Terminals on the PCB were soldered to 31-gauge wires and subsequently connected to shielded coaxial cables from which the accelerometer output voltage was measured.
The output voltage was AC coupled, amplified, and sampled at 100~kHz using PowerLab ML880 (ADInstruments). 
The thin, flexible wires attached to the PCB were meant to reduce the inertial effects of the coaxial cables on the accelerometer.
The PCB was placed on the flat antero-medial face of the tibial shaft 16~cm proximal to the tapper such that the accelerometer was directly in contact with the skin.
It was secured to skin using adhesive gauze tape (BSN Cover-Roll Stretch, Hamburg, Germany).

\subsection*{Spectral analysis}

All signal processing was performed in Python v3.11.5.
The raw acceleration signal was segmented into 2~ms long segments starting at 0.15~ms after each tap.
There was insignificant energy in the signal beyond 2~ms.
The signals for participant 5 were truncated at 1~ms due to noise in the later portions.    
Each segment was mean-centered and normalized by the maximum acceleration and windowed using a cosine window of the same length.
The windowed signals were transformed into the frequency domain using the Fast Fourier Transform (FFT) with 40000 bins (\texttt{NumPy v1.26.4}).
The large FFT size resulted in efficient sinc interpolation of the spectrum.
The magnitudes of the acceleration spectra resulting from each tap were arranged into a spectrogram with frequency and time on orthogonal axes (example in figure~\ref{fig:results-static}A), such that a vertical slice of the spectrogram at a time instant contained the acceleration spectrum of the wave resulting from a tap at that instant.
We used a generic peak finding function with default parameters (\texttt{find\_peaks} in \texttt{SciPy v1.12.0}) to find locations of the local maxima in the spectrum at each time instant.
To prevent abrupt jumps in the spectral peak locations from one time instant to the next, we constrained the peak finder to only search between specified frequency limits and tapered the spectrogram with a first-order Butterworth bandpass filter spanning these limits.
The outcome of this process was a time-frequency ridge of peaks spanning the entire duration of the spectrogram at 50~Hz (example ridge shown by black curve in figure~\ref{fig:results-static}A).

In general, the spectrograms showed multiple ridges in different frequency windows, and these windows differed from trial to trial (figure~\ref{fig:swaying-data-supplement}).
Using a calibration step, we determined a region of interest (ROI), \emph{i.e.} the frequency window in the spectrogram that contained peak locations that best correlated with tibial load (details in SI section~\ref{sec:calibration details}).
Calibration was performed using the first 6~s of data from each trial.
This subset of data contained enough variance to be representative of the entire trial and longer calibration lengths did not meaningfully improve the results.
In this subset spectrogram, the peak finder extracted peak ridges within many candidate windows.
The extracted peak ridge from each candidate window was compared with the measured reference tibial compressive force in the calibration subset.
The frequency window that yielded the ridge with the highest correlation with tibial load was chosen as the ROI.
Once calibration was completed, the initial 6~s subset of force and frequency data were excluded from any subsequent goodness-of-fit tests.
In the remainder of that trial (unseen during calibration), the peak finder was restricted to the calibrated ROI and the peak ridge extracted from it was used in subsequent analyses. 

\subsection*{Statistical analyses}
For each swaying trial, the Pearson correlation coefficient was calculated between the reference tibial compressive force (downsampled to 50~Hz) and the spectral peak ridge over the entirety of the trial excluding the calibration subset.
For walking, the data were segmented into stance and swing phases excluding the calibration subset. A 20~N vertical ground reaction force threshold was used to identify foot contact events and the data were normalized to lie between 0\% (heel strike) and 100\% (toe off) of stance. Pearson correlation coefficients between the reference tibial compressive force and the spectral peak ridge were calculated for each stance phase and the mean correlation coefficient across all stance phases was calculated for each participant.
Correlation coefficients from each stance phase were first transformed to Fisher z-scores, averaged, and the average z-score was transformed back to a correlation coefficient using the inverse Fisher transform \citep{corey1998averaging,stetter2020machine,matijevich2019ground}. 
Mean correlation coefficients across all participants were also calculated using the same method.

\section*{Acknowledgements}
We thank V. Joshi, K. Nguyen, A. Ravan, E. Reimink, and N. Sharma for helpful discussions and feedback at various stages of this study. All authors acknowledge support from the American School of Prehistoric Research (Harvard University).

\bibliographystyle{abbrvnat}

\clearpage
\appendix
\renewcommand{\theequation}{S\arabic{equation}}
\renewcommand{\thefigure}{S\arabic{figure}}
\renewcommand{\thesection}{S\arabic{section}}
\renewcommand{\thesubfigure}{\roman{subfigure}}
\setcounter{equation}{0}
\setcounter{figure}{0}
\input{supplement}

\end{document}

%% file: supplement.tex
% MACROS
\def\fig{figure}
\def\tab{table}
\def\eqn{equation}

\date{}

\section*{\centering \LARGE\bfseries Supplementary Information}
\vspace{1cm}

\renewcommand{\theequation}{S\arabic{equation}}
\renewcommand{\thefigure}{S\arabic{figure}}
\renewcommand{\thesection}{S\arabic{section}}
\renewcommand{\thesubfigure}{\roman{subfigure}}
\setcounter{equation}{0}
\setcounter{figure}{0}

\section{Axially compressed Euler-Bernoulli beam model of the tibia}
\subsection{General far-field solution for propagating flexural waves}
\label{sec:EB derivation}
In this section, we derive expressions for the spectra of propagating flexural waves in the axially compressed tibia, following the beam modelling approach in \citet{doyle2020wave} and \citet{graff2012wave}.
We model the tibia as an infinitely long Euler-Bernoulli beam with uniform density $\rho$, cross-sectional moment of inertia $I$, Young's modulus $E$, and cross-sectional area $A$. A uniform compressive force $P$ (interchangeably called ``axial load'') is applied along the long axis of the tibia. The governing partial differential equation is

\begin{equation}
    EI\frac{\partial^4 {y}}{\partial {x}^4} + {P}\frac{\partial^2 {y}}{\partial {x}^2} + \rho A\frac{\partial^2 {y}}{\partial {t}^2} = 0, 
    \label{eq:EB equation dimensional}
\end{equation}
where $x$ is the axial coordinate, $t$ is time, $y(x,t)$ is the transverse displacement. With the radius of gyration $r=\sqrt{I/A}$ as the length scale, the speed of sound in the medium $c=\sqrt{E/\rho}$ to define the time scale $r/c$, and axial rigidity $EA$ as the force scale, we non-dimensionalize the variables in equation~\eqref{eq:EB equation dimensional} using the following definitions
\begin{equation*}
    \hat{x} \equiv \dfrac{x}{r},\quad \hat{y} \equiv \dfrac{y}{r}, \quad \hat{t} \equiv \dfrac{t c}{r}, \quad \hat{P} \equiv \dfrac{P}{EA}.
\end{equation*}
The resulting dimensionless Euler-Bernoulli equation is given by
\begin{equation}
    \frac{\partial^4 \hat{y}}{\partial \hat{x}^4} + \hat{P}\frac{\partial^2 \hat{y}}{\partial \hat{x}^2} + \frac{\partial^2 \hat{y}}{\partial \hat{t}^2} = 0, 
    \label{eq:EB equation dimensionless}
\end{equation}
We solve equation~\eqref{eq:EB equation dimensionless} in the frequency domain to obtain a solution of the form $\mathcal{Y}(\hat{x},\hat{\omega})$ that specifies the frequency spectrum of the beam's transverse displacement at a location $\hat{x}$ on the beam. 
This spectrum is defined via the Fourier transform as $\mathcal{\hat{Y}}(\hat{x},\hat{\omega}) = \int_{-\infty}^{\infty} \hat{y}(\hat{x},\hat{t})e^{-i\hat{\omega}\hat{t}}d\hat{t}$.
Applying the Fourier transform to equation \eqref{eq:EB equation dimensionless} we get the spectral form of the Euler-Bernoulli equation,
\begin{equation*}
    \frac{d^4 \mathcal{\hat{Y}}}{d \hat{x}^4} + \hat{P}\frac{d^2 \mathcal{\hat{Y}}}{d \hat{x}^2} - \hat{\omega}^2\mathcal{\hat{Y}} = 0.
\end{equation*}
The general solution to this ordinary differential equation is given by $\mathcal{Y} = C(\hat{k})~e^{\hat{k}\hat{x}}$, where $\hat{k}$ is the complex, frequency-dependent wavenumber satisfying the characteristic equation
\begin{equation*}
    \hat{k}^4 + \hat{P}\hat{k}^2 - \hat{\omega}^2 = 0.
\end{equation*}
There are two real and two imaginary solutions for the wavenumber given by $\hat{k}=\pm i\hat{\xi}, \pm\hat{\kappa}$ where
\begin{equation*}
    \hat{\xi}(\hat{P},\hat{\omega}) = \sqrt{\dfrac{\hat{P}+\sqrt{\hat{P}^2 + 4\hat{\omega}^2}}{2}}, \quad\quad
    \hat{\kappa}(\hat{P},\hat{\omega}) = \sqrt{\dfrac{-\hat{P}+\sqrt{\hat{P}^2 + 4\hat{\omega}^2}}{2}}, \quad\quad \quad\quad
\end{equation*}
and $\hat{\xi},\hat{\kappa}\in\mathbb{R}^+$. 
Rewriting these in dimensionful form yields the expressions in equation~\eqref{eq:wavenumbers} in the main text.

We can then write the general solution in the frequency domain as $\mathcal{\hat{Y}} = C_1e^{-i\hat\xi \hat{x}}+C_2e^{-\hat\kappa \hat{x}}+C_3e^{i\hat\xi \hat{x}}+C_4e^{\hat\kappa \hat{x}}$,
where $C_n$ are complex-valued functions of the wavenumbers. The general solution consists of a forward (along positive $\hat{x}$) and a backward traveling wave, along with spatially decaying (``near-field'') terms. We restrict our analysis to solutions for $\hat{x} \geq 0$ given by,
\begin{equation}
    \mathcal{\hat{Y}} = C_1e^{-i\hat{\xi} \hat{x}}+C_2e^{-\hat{\kappa} \hat{x}}.
    \label{wave solutions}
\end{equation}

\subsection{Boundary conditions (infinite domain)}
\label{sec:boundary conditions}
We require two boundary conditions to solve for the coefficients $C_1, C_2$. For the first boundary condition, we assume that the resulting waves travel symmetrically outward from the point of tapping giving us the boundary condition $d\mathcal{\hat{Y}}/d\hat{x} = 0$ at $\hat{x}=0$.
Applying this boundary condition to equation~\eqref{wave solutions} we get $C_2 = -i(\hat\xi/\hat\kappa) C_1$, so that the solution can be written with a single unknown coefficient as,
\begin{equation}
    \mathcal{\hat{Y}} = C_1(e^{-i\hat\xi \hat{x}} -i\dfrac{\hat\xi}{\hat\kappa}e^{-\hat\kappa \hat{x}}).
    \label{eq:single unknown P}
\end{equation}

We solve for $C_1$ using shear force balance at the location of tapping, $\hat{x}=0$ where the tapper applies a force $\hat{f}(\hat{t})$. By symmetry, we only need to consider the right half of the beam, supporting half of the applied force. We assume that this force is applied to a massless particle of the beam for which the force balance is given by $\hat{f}/2 + \hat{v}_r = 0$, where $\hat{v}_r=-{d^3 \mathcal{\hat{Y}}}/{d\hat{x}^3}$ is the shear force applied from the right half of the beam to the particle (figure~\ref{fig:theory}A). If the Fourier transform of the applied force $\hat{f}(\hat{t}$) is $\mathcal{\hat F}(\hat{\omega})$, we can write
\begin{equation}
    \dfrac{\mathcal{\hat{F}}}{2} - \left .\dfrac{d^3\mathcal{\hat{Y}}}{d\hat{x}^3}\right |_{\hat{x}=0} = 0 \implies
    C_1  = \dfrac{-i\mathcal{\hat{F}}}{2(\hat\xi^3 + \hat\xi\hat\kappa^2)}.
    \label{eq:solving for P}
\end{equation}
From equations \eqref{eq:single unknown P} and \eqref{eq:solving for P}, the complete frequency domain solution for the flexural wave displacement is given by
\begin{equation*}
    \mathcal{\hat{Y}} = \dfrac{-i \mathcal{\hat{F}}}{2(\hat\xi^3 + \hat\xi\hat\kappa^2)} (e^{-i\hat\xi \hat{x}} -i\dfrac{\hat\xi}{\hat\kappa}e^{-\hat\kappa \hat{x}}).
\end{equation*}
% The displacement is a function of $\hat{\omega}, \hat{P}$ and $\hat{x}$.
Far from the location of tapping ($\hat{x}\gg 0$), the spatially decaying (``near-field'') term vanishes and the solution is simplified to the far-field limit
\begin{equation}
    \mathcal{\hat{Y}} = \dfrac{-i\mathcal{\hat{F}}}{2(\hat\xi^3 + \hat\xi\hat\kappa^2)}e^{-i\hat\xi \hat{x}}.
    \label{eq:far field solution}
\end{equation}
Here, we define the transfer impedance of the beam as the ratio of the force at the origin and velocity at a far-field location $\hat{x}_m \gg 0$, $\mathcal{\hat Z} \equiv \mathcal{\hat F}/{i\hat\omega\mathcal{\hat Y}}$. Its magnitude is given by
\begin{equation}
\hat{Z} = \dfrac{2(\hat\xi^3 + \hat\xi\hat\kappa^2)}{\hat\omega}.
\label{eq:impedance expression}
\end{equation}
The dimensionful version of this equation yields equation~\eqref{eq:impedance} in the main text.
Finally, the magnitude of the acceleration spectrum, $\hat{a}$, may be expressed using the magnitude of the transfer impedance as 
\begin{equation*}
    \hat a(\hat{P},\hat{\omega}) = \dfrac{\hat{\omega}|\mathcal{\hat F}|}{{\hat{Z}(\hat{P},\hat{\omega})}}.
    \label{eq:impedance definition}
\end{equation*}

\subsection{Load-dependence of peaks in the acceleration magnitude spectrum}
\label{sec:peak shift derivation}

As axial load increases, the impedance selectively suppresses lower flexural wave frequencies (figure~\ref{fig:theory}B).
One measurable consequence of this phenomenon is that any peaks in the acceleration spectrum would shift to higher frequencies under higher loads.
To derive this relationship, we examine the acceleration spectra of the loaded and zero-load beams. 
Normalizing the acceleration spectrum of the loaded beam by the spectrum of the zero-load beam we get
\begin{equation}
    \dfrac{{\hat{a}}(\hat{P},\hat{\omega})}{{\hat{a}}_0(\hat{\omega})} = \dfrac{\hat Z(0,\hat{\omega})}{\hat Z(\hat{P},\hat{\omega})} \equiv M(\hat{P},\hat{\omega}),
    \label{eq:Z ratio}
\end{equation}
where $\hat{a}_0(\hat{\omega}) \equiv \hat{a}(0,{\hat\omega})$. 
At $\hat{P}=0$, we assume that the magnitude of the acceleration spectrum has a peak located at $\hat{\omega}^*$, so that $\hat{a}'_0(\hat{\omega}^*)=0$, where prime denotes derivative with respect to $\hat{\omega}$. 

We note that for the human tibia, Young's modulus $E\approx20$~GPa and cross-sectional area at the tibial mid-shaft $A\approx2.5 \times 10^{-4}$~m$^2$, resulting in an axial rigidity of about $5\times10^{6}$~N. 
So, for tibial loads seen in our data (approximately 1000 N) the dimensionless compressive force on the tibia $\hat{P} \sim 10^{-4} \ll 1$.
Thus, linear approximations around $\hat{P} = 0$ are justified.  

Under a small compressive force $\Delta\hat{P}\ll 1$, let the peak of the acceleration magnitude shift from $\hat{\omega}^*$ to a nearby frequency $\hat{\omega}^* + \Delta\hat{\omega}$, 
so that $\hat{a}'(\Delta\hat{P},\hat{\omega}^* + \Delta\hat{\omega})= 0$. 
Differentiating $\hat{a}(\hat{P},\hat\omega)$ with respect to $\hat{\omega}$ in equation~\eqref{eq:Z ratio} we can write
\begin{equation}
     M'(\Delta\hat{P},\hat{\omega}^* +\Delta\hat{\omega}) \hat{a}_0(\hat{\omega}^* +\Delta\hat{\omega}) + M(\Delta\hat{P},\hat{\omega}^* +\Delta\hat{\omega})\hat{a}'_0(\hat{\omega}^* +\Delta\hat{\omega}) = 0.
    \label{eq:at derivative}
\end{equation}
To evaluate equation \eqref{eq:at derivative}, we use linear approximations of $\hat{a}_0,\hat{a}'_0$ around $\hat{\omega}=\hat{\omega}^*$, and of $M, M'$ around $\hat{P}=0$ and $\hat{\omega}=\hat{\omega}^*$:
\begin{align}
    \begin{split}
    \hat{a}_0(\hat\omega^*+\Delta\hat\omega) &\approx \hat{a}_0(\hat\omega^*), 
    \label{eq:taylor1}
    \\
    \hat{a}'_0(\hat\omega^*+\Delta\hat\omega) &\approx \hat{a}''_0(\hat\omega^*)\Delta\hat\omega, 
    % \label{eq:taylor2}
    \\
    M(\Delta\hat{P},\hat\omega^* + \Delta\hat\omega) &\approx M(0,\hat\omega^*) + \left . \Delta\hat{P}\dfrac{\partial M}{\partial \hat{P}}\right |_{0,\hat\omega^*} + \left .\Delta\hat\omega\dfrac{\partial M}{\partial \hat\omega}\right |_{0,\hat\omega^*} = 1 - \dfrac{\Delta\hat{P}}{4\hat\omega^*},
    % \label{eq:taylor3}
    \\
    M'(\Delta\hat{P},\hat\omega^* + \Delta\hat\omega) &\approx M'(0,\hat\omega^*) + \left . \Delta\hat{P}\dfrac{\partial M'}{\partial \hat{P}}\right |_{0,\hat\omega^*} + \left .\Delta\hat\omega\dfrac{\partial M'}{\partial \hat\omega}\right |_{0,\hat\omega^*} = \dfrac{\Delta\hat{P}}{4{\hat\omega^{*2}}}.
    % \label{eq:taylor4}
    \end{split}
\end{align}
Using equations \eqref{eq:taylor1} to evaluate equation \eqref{eq:at derivative}, we can write,
\begin{equation*}
    \hat a_0(\hat\omega^*)\dfrac{\Delta\hat{P}}{4{\hat\omega^{*2}}} + \hat a''_0(\hat\omega^*)\Delta\hat\omega\left (1 - \dfrac{\Delta\hat{P}}{4{\hat\omega^*}}\right ) = 0.
\end{equation*}
It follows that
\begin{equation*}
    \Delta\hat\omega = -\dfrac{\hat{a}_0(\hat\omega^*)}{\hat a''_0(\hat\omega^*)}\dfrac{\Delta\hat{P}}{4{\hat\omega^{*2}}}\left (1 - \dfrac{\Delta\hat{P}}{4{\hat\omega^*}}\right )^{-1}.
\end{equation*}
For $x \ll 1$, we have $(1-x)^{-1}\approx 1 + x$.  Ignoring the term with $\Delta P^2$, we can write,
\begin{equation*}
    \Delta\hat\omega = -\dfrac{\hat{a}_0(\hat\omega^*)}{\hat a''_0(\hat\omega^*)}\dfrac{\Delta\hat{P}}{4{\hat\omega^{*2}}}.
    \label{eq:delta omega}
\end{equation*}
Rewritten in dimensionful variables and noting that the at a peak, $a''_0(\omega^*) < 0$, we see that spectral peak shifts are related to axial load as
\begin{equation}
    \Delta\omega = \dfrac{a_0(\omega^*)}{|a''_0(\omega^*)|}\dfrac{\Delta P}{{\displaystyle 4\sqrt{\params}}{\omega^{*2}}}.
    \label{eq:delta omega dimensionful}
\end{equation}
For brevity, and since the linearization is done around $P=0$, we replace $\Delta P$ with $P$ in the main text.
Equation~\eqref{eq:delta omega dimensionful} yields several insights. First, the higher the peak frequency $\omega^*$ at zero load, the smaller the change in peak frequency. 
This is because at higher frequencies, the impedance of the axially compressed beam approaches that of the zero-load beam. 
Second, the ratio $a_0/|a''_0|$ (same as $\gamma$ in equation~\eqref{eq:main result linear}) quantifies the sharpness of the spectral peak at zero load. 
This ratio is governed only by the tapping force $\mathcal{F}(\omega)$ and is independent of the beam's material properties. 
The sharper the zero-load peak, the smaller the ratio and thus smaller the change in peak frequency. 
In the limit of a zero-load peak that approaches a delta-function, changes in the impedance would only scale the height of the peak and not shift the peak.  

\subsection{Load-dependence of the acceleration phase spectrum}

From equation~\eqref{eq:far field solution}, we see that the phase spectrum of the acceleration also depends on the tibial compressive force. The phase is defined as $\phi(\omega) \equiv \arg\mathcal{\hat{A}} = -\hat{\xi}\hat{x}_m$.
Normalizing the phase for a loaded beam by the phase of the zero-load beam and linearizing around $\hat{P}=0$ we get,
\begin{equation}
    \dfrac{\phi}{\phi_0} = \left (1 + \dfrac{\Delta\hat{P}}{4\hat\omega}\right )  \equiv 1 + \dfrac{\Delta P}{4\omega\displaystyle\sqrt{\params}}.
    \label{suppeq: phase}
\end{equation}
Thus, at any frequency, the phase difference measured for a propagating flexural wave has an affine dependence on the compressive force.

\section{Validity and impacts of modeling assumptions}
\label{sec:model validity}
\subsection{Euler-Bernoulli beam model}

The Euler-Bernoulli beam model is applicable for thin, slender beams, where the wavelength of propagating waves is much larger than the thickness of the beam \citep{doyle2020wave}. Experimental data from \citet{vogl2016reliability} show that tibial flexural waves of frequency 3000~Hz have a phase velocity of about 480 m/s, which results in a wavelength of about 16~cm. Since flexural wave phase velocity scales with the square root of frequency \citep{doyle2020wave}, at lower frequencies the wavelengths are even larger. Thus, for the frequency range of interest to us, the flexural wavelength is much larger than the typical diameter of the tibial mid-shaft.

\subsection{Uniform material and geometric properties}

Our model assumes that material properties (density, Young's modulus) and geometric properties (cross-sectional area and moment of inertia) are uniform through the length of the beam.
These properties are also assumed to be independent of the axial compressive force.
Inhomogeneities in the tibia undoubtedly affect wave propagation, but we anticipate the effects to be minimal for two reasons.
First, experimental measurements of tibial resonant frequencies have been found to be comparable to predictions from simple, uniform beam models (5.8\% error, \citet{collier1982mechanical}). 
In their study on \emph{in vivo} flexural wave phase velocity measurements in the tibia, \citet{stussi1988assessment} also find that a homogenous Euler-Bernoulli model is sufficient to capture the essential flexural dynamics of the tibia.
Second, since the tapper and the accelerometer are in fixed locations on the tibia throughout a trial, the effects of any spatial inhomogeneities are common to all flexural waves measured in the trial.
As we only measure changes in the spectrum (specifically, shifts in spectral peak locations), these effects may be ignored.

\subsection{Infinite vs. finite beam length}

Experimentally, a finite beam may be considered infinite for analysis if wave reflections from the boundaries are negligible. 
This is possible if we can guarantee that the time window of the measurement precisely excludes any reflections, or if damping in the beam is so high that the reflections have negligible power by the time they arrive at the measurement location.
Both these assumptions are partially met in our setup.
The fastest tibial flexural wave at 3000 Hz travels at around 480 m/s \citep{vogl2016reliability}, which suggests that the round trip travel time between the accelerometer and a boundary is just under 1~ms for a tibia length of about 40 cm. Our measurement window of 2~ms (chosen to capture most of the response signal's energy) is thus long enough to allow one or two of the fastest traveling reflections to arrive at the measurement site. We also see that the signal is dampened to almost zero energy at the end of this short window (sample in figure 2C), suggesting that reflections are largely attenuated by the time they arrive at the accelerometer. Thus, the infinite length assumption is a reasonable approximation. 

\begin{figure}[t]
\centering
\includegraphics[width=0.5\textwidth]{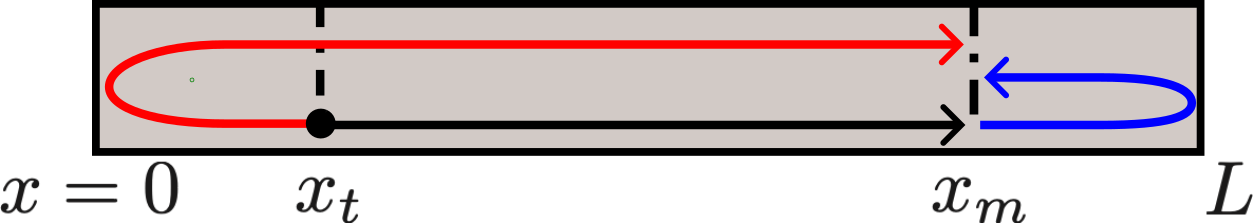}
\caption{\textbf{Flexural waves in a finite beam.} A single ``tap'' at $x_t$ generates two symmetric wave pulses that travel outward from the source. The directions and reflections of the pulses are shown schematically by colored paths. The right-going pulse travels along the black path and directly reaches the accelerometer at $x_m$. The pulse goes past the accelerometer along the blue path and gets reflected at the boundary at $x=L$ before arriving at $x_m$. The initially left-going pulse along the red path propagates to the left, gets reflected at the boundary at $x=0$ and arrives at $x_m$.  The resulting measurement by the accelerometer at $x_m$ is the sum of these three pulse arrivals.
\label{fig:reflection-schematic}}
\end{figure}

This assumption greatly simplifies the theoretical analysis of wave propagation and is often the starting point in the study of elastic beams \citep{doyle2020wave}.
The solution for an infinite beam yields the baseline shape of the propagating wave's spectrum and its dependence on the beam's material properties, cross-sectional geometry, and axial load. 
This baseline spectrum is modulated by reflections from the boundaries when the beam's finite extent is considered.
Consider a finite beam with endpoints at $x=0,L$. The source of flexural waves is a tapper located at $x_t$ and the wave is measured by an accelerometer at $x_m$. A single tap generates two symmetric pulses emanating in the left and right directions (red and black paths respectively, figure~\ref{fig:reflection-schematic}), ultimately reaching the accelerometer. If we can guarantee that only the directly incident, right traveling wave is measured by the accelerometer (black path, figure~\ref{fig:reflection-schematic}), then the infinite beam assumption is exactly valid. Starting from the expression for a far-field propagating wave from equation \eqref{eq:far field solution}, the measured acceleration spectrum in this case may be expressed as
\begin{equation*}
a_{\infty}(\omega) = G(\omega) e^{-i\xi (x_m - x_t)}.
\end{equation*}
Here, $|G(\omega)|$ is the magnitude of the infinite beam's acceleration spectrum.
We now consider reflections arriving at the accelerometer. 
For this analysis, only the first reflections from each boundary are considered. 
The boundaries at the two ends have reflection coefficients $R_0$ and $R_L$ which are in general complex and depend on frequency and the type of boundary condition (fixed, pinned, etc.). 
The right traveling pulse travels past the accelerometer to the boundary at $x=L$ before being reflected back to the accelerometer (blue path, figure~\ref{fig:reflection-schematic}). This reflected wave travels a total path length $2L - x_m - x_t$ and has the spectrum
\begin{align*}
a_{L}(\omega) &= R_LG(\omega)e^{-i\xi(2L - x_m - x_t)}.
\end{align*}
The initially left-going pulse (red path, figure~\ref{fig:reflection-schematic}) reflects at the boundary at $x=0$ and travels the path length $x_t+x_m$ before arriving at the accelerometer with the spectrum
\begin{align*}
a_0(\omega) &= R_0G(\omega)e^{-i\xi(x_t + x_m)}.
\end{align*}
The spectrum of the total signal is given by the sum of the three arrivals and can be rearranged as
\begin{equation*}
    a(\omega) = G(\omega) e^{-i\xi (x_m - x_t)}\left (1 + R_L e^{-i2\xi (L-x_m)} + R_0 e^{-i2\xi x_t}\right).
    \label{eq:reflection-spectrum}
\end{equation*}
The magnitude of this spectrum is 
\begin{equation}
    |a(\omega)| = |G(\omega)|\ |1 + R_L e^{-i2\xi (L-x_m)} + R_0 e^{-i2\xi x_t} |,
    \label{eq:reflection-spectrum-magnitude}
\end{equation}
which comprises the infinite beam's response $|G(\omega)|$ and the contribution from reflections.
Thus, the overall shape of the finite beam's response captures the infinite beam's response, which undergoes a sinusoidal modulation depending on which frequencies in the three arrivals are in and out of phase.
Since the boundaries of the tibia are unlikely to be perfect reflectors, $|R_L| < 1$ and  $|R_0| < 1$. 
Further, since there is damping in the system the strength of the each pulse decreases with the path length it travels before arriving at the accelerometer.
In such a situation the last two terms in equation~\eqref{eq:reflection-spectrum-magnitude} are small, so the shape of the finite beam's acceleration spectrum is a good approximation of the infinite beam spectrum, which we know is load-dependent.

If more than two reflections are measured, additional terms are added to the right hand side of equation~\eqref{eq:reflection-spectrum-magnitude} with increasing path lengths and higher powers of the reflection coefficients.
In the extreme limit of zero damping, perfect reflectors, and a long measurement window, there are an infinite number of reflections from both boundaries, resulting in a sampling of the spectrum $|G(\omega)|$ such that it remains nonzero only at resonant frequencies of the beam. In this limit the waves are no longer propagating waves but instead become standing waves whose resonant frequencies decrease as compressive force increases \citep{bokaian1988natural}.

\section{Details of the sensor calibration process}
\label{sec:calibration details}
The Euler-Bernoulli beam model predicts that the entire spectrum of flexural waves shifts towards higher frequencies with increasing tibial load.
In reality, the spectrum likely contains contributions from vibrations that are not flexural \citep{stussi1988assessment,fah1988phase}, as well as noise that is trial dependent. 
As a consequence, the structure of acceleration spectra varies widely across participants (figure~\ref{fig:swaying-data-supplement}), and not all peaks in the spectra are necessarily sensitive to tibial load.

To mitigate this variability, we performed a calibration step for each trial, whose goal was to identify a frequency region of interest (ROI) in the spectrogram, \emph{i.e.} the range of frequencies which contained peaks that most sensitively varied with the reference tibial compressive force.
Data from an initial 6~s subset of each trial were used to perform the calibration.
In the subset spectrogram, we used the peak finder to find peaks in various candidate frequency windows defined by the bounds ($f_{low}, f_{high}$).
The candidate windows were chosen to span the entire range of frequencies seen in our data (0 to 3400~Hz), with each window having a minimum bandwidth of 200~Hz.
To accomplish this, the lower bound $f_{low}$ took on values from 0 to 3200 Hz in increments of 200~Hz. The upper bound $f_{high}$ then took on values $f_{low} + 200n$ where $n = 1,2,3...$, with $f_{high}$ capped at 3400~Hz.
Further increases in the candidate window resolution did not meaningfully improve the results. 
For each window, the spectrogram was tapered with a first-order Butterworth bandpass filter spanning the window limits ($f_{low}$ was set to 0.1~Hz instead of 0~Hz to get a valid bandpass filter).
Subsequently, the peak finder was applied to the subset spectrogram and a ridge of peak locations was extracted from each candidate window.
A total of 153 candidate windows were considered and the ridge extracted from each window was correlated with the tibial load in the calibration subset.
The window that contained the ridge with the highest correlation with tibial load was chosen as the ROI.
When necessary, the ROI was adjusted to ensure that peaks were not clipped.
In the remainder of the trial, the peak finder was restricted to search within this ROI.
Data used in calibration were removed from any subsequent goodness-of-fit calculations, so all reported results are based on data unseen during calibration. Values of $f_{low}$ and $f_{high}$ identified in each trial are provided in the electronic supplementary data. 

\bibliographystyle{abbrvnat}

\newpage
\subsection*{Additional data figures}
\begin{figure}[h!]
    \centering
    \begin{subfigure}[b]{0.4\textwidth}
    \includegraphics[width=\textwidth]{figures/swayplots/results-sway-S01.pdf}
    \caption*{Participant 1.}
    \end{subfigure}
    \hspace{3pt}
    \begin{subfigure}[b]{0.4\textwidth}
    \includegraphics[width=\textwidth]{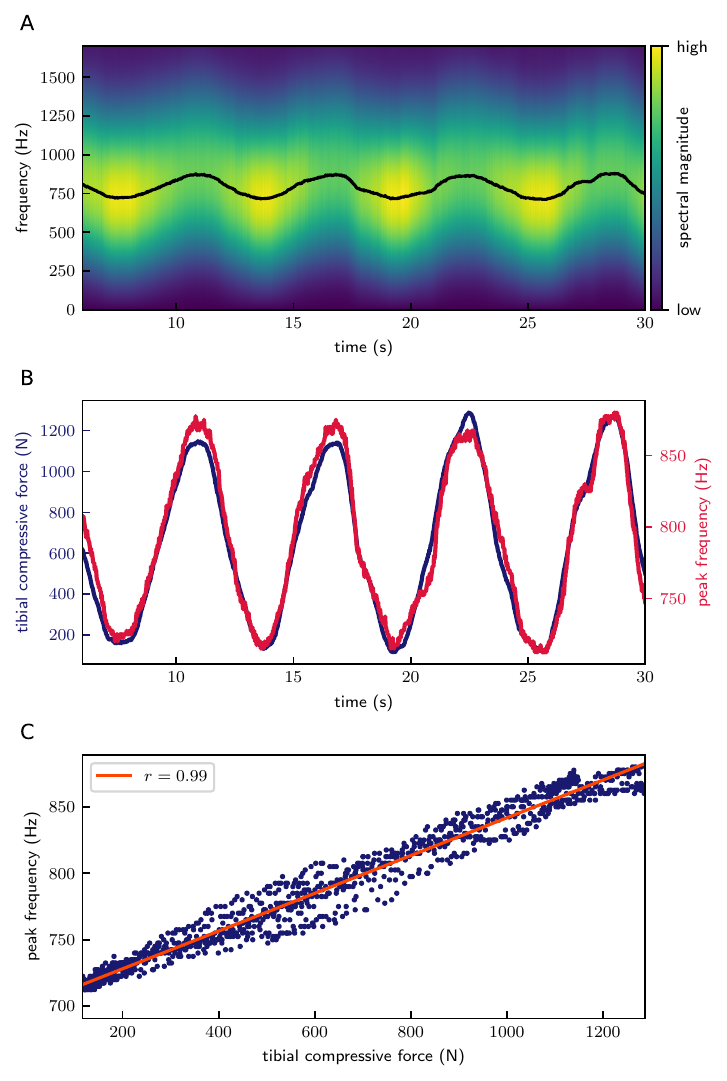}
    \caption*{Participant 2.}
    \end{subfigure}
    \centering
    \begin{subfigure}[b]{0.4\textwidth}
    \includegraphics[width=\textwidth]{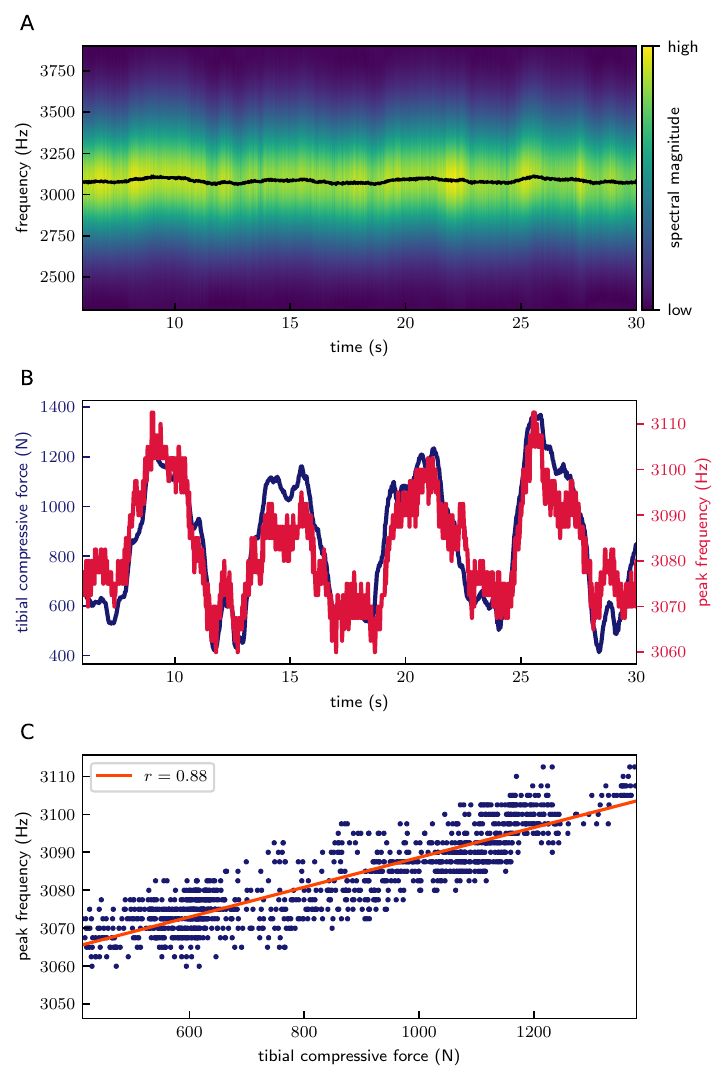}
    \caption*{Participant 3.}
    \end{subfigure}
    \hspace{3pt}
    \centering
    \begin{subfigure}[b]{0.4\textwidth}
    \includegraphics[width=\textwidth]{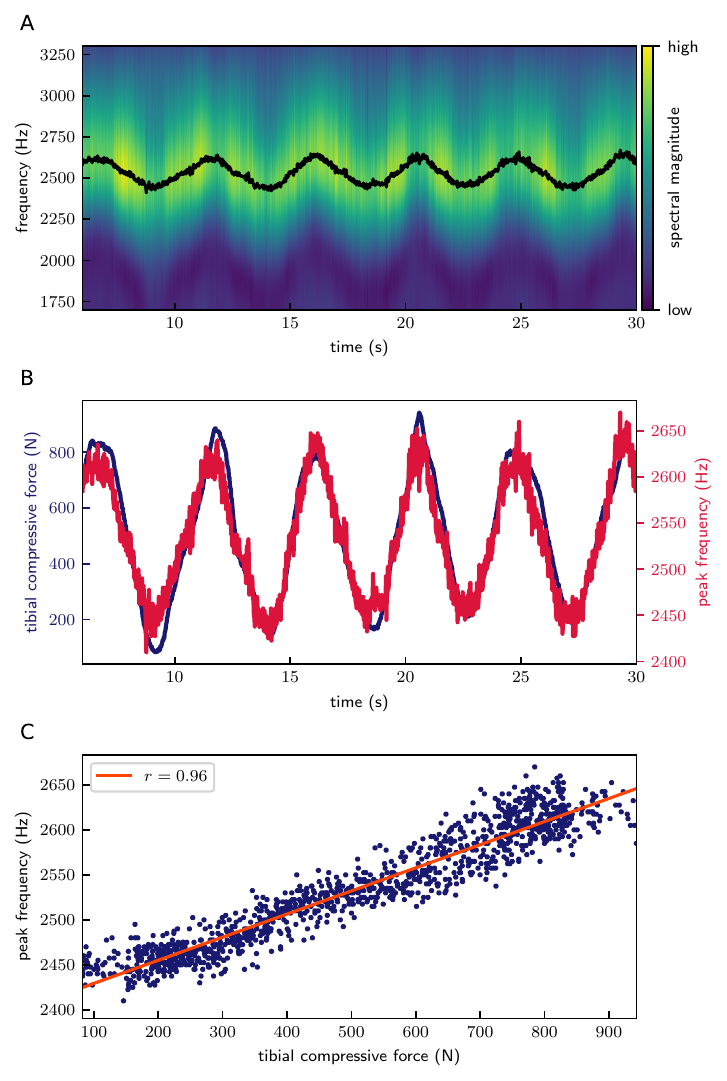}
    \caption*{Participant 4.}
    \end{subfigure}
    \caption{Data from swaying trials for all participants}
    \label{fig:swaying-data-supplement}
\end{figure}

\begin{figure}[h!]
    \ContinuedFloat
    \centering
    \begin{subfigure}[b]{0.4\textwidth}
    \includegraphics[width=\textwidth]{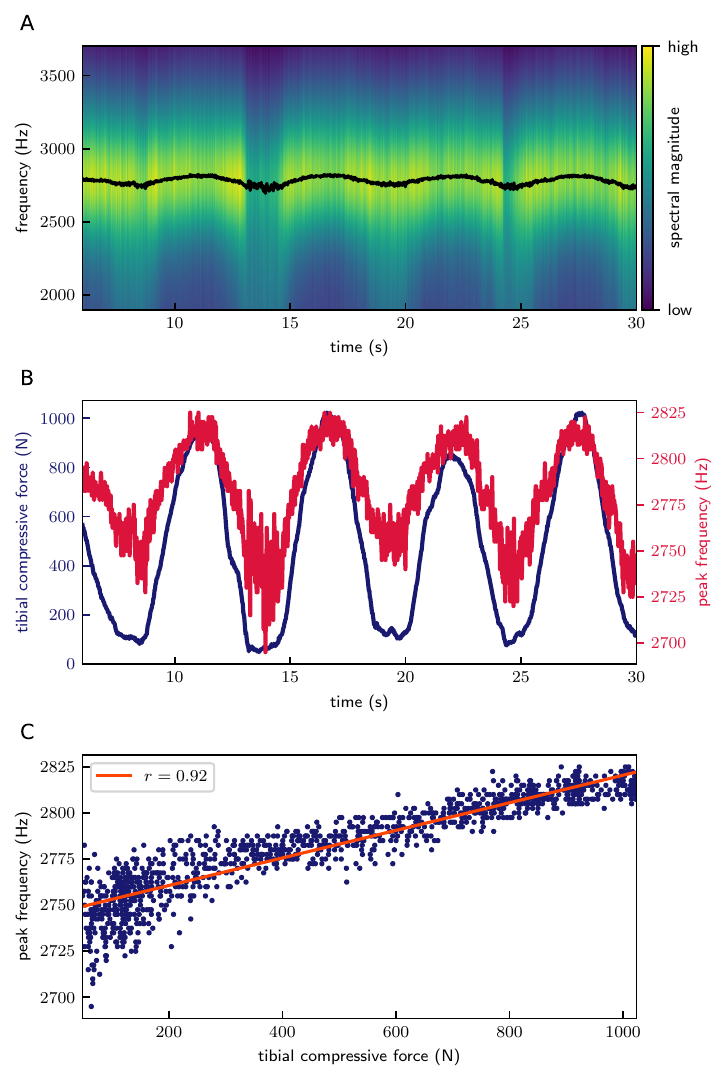}
    \caption*{Participant 5.}
    \end{subfigure}
    \hspace{3pt}
    \begin{subfigure}[b]{0.4\textwidth}
    \includegraphics[width=\textwidth]{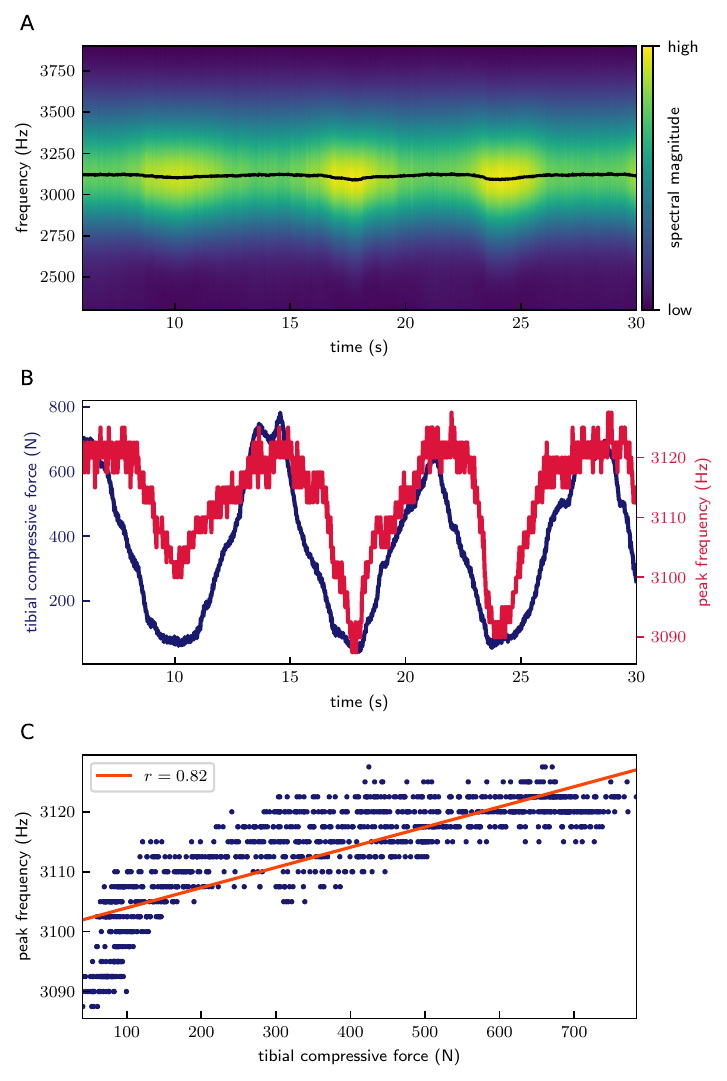}
    \caption*{Participant 6.}
    \end{subfigure}
    \centering
    \begin{subfigure}[b]{0.4\textwidth}
    \includegraphics[width=\textwidth]{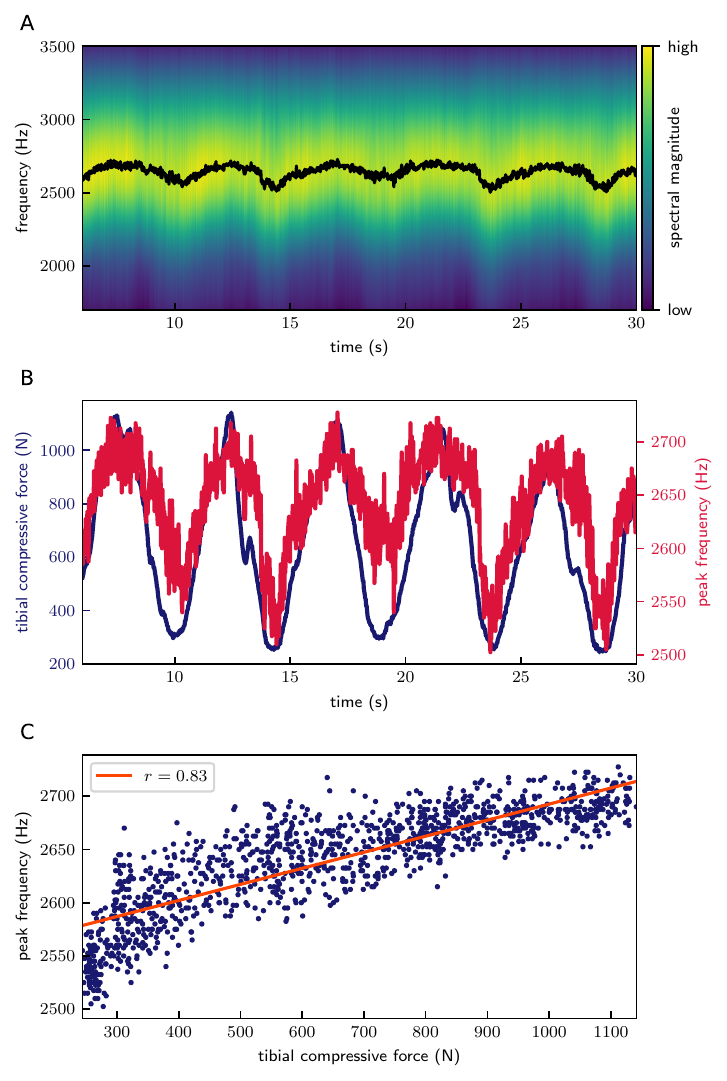}
    \caption*{Participant 7.}
    \end{subfigure}
    \hspace{3pt}
    \centering
    \begin{subfigure}[b]{0.4\textwidth}
    \includegraphics[width=\textwidth]{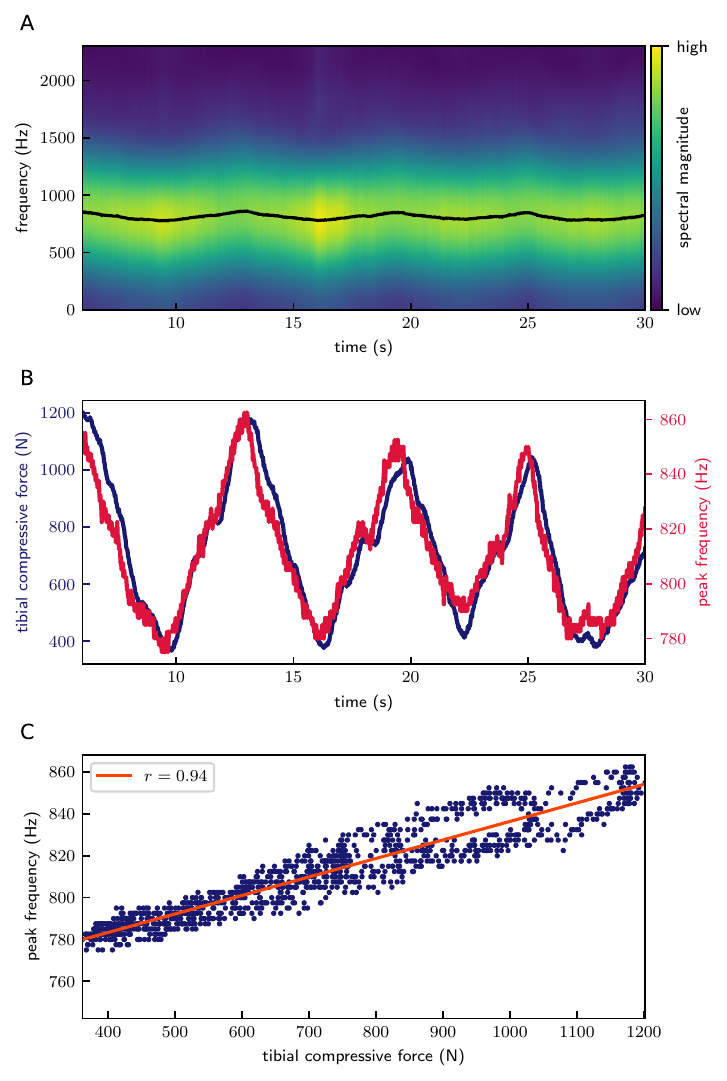}
    \caption*{Participant 8.}
    \end{subfigure}
    \caption{Data from swaying trials for all participants}
\end{figure}

\begin{figure}[t!]
    \ContinuedFloat
    \centering
    \begin{subfigure}[b]{0.4\textwidth}
    \includegraphics[width=\textwidth]{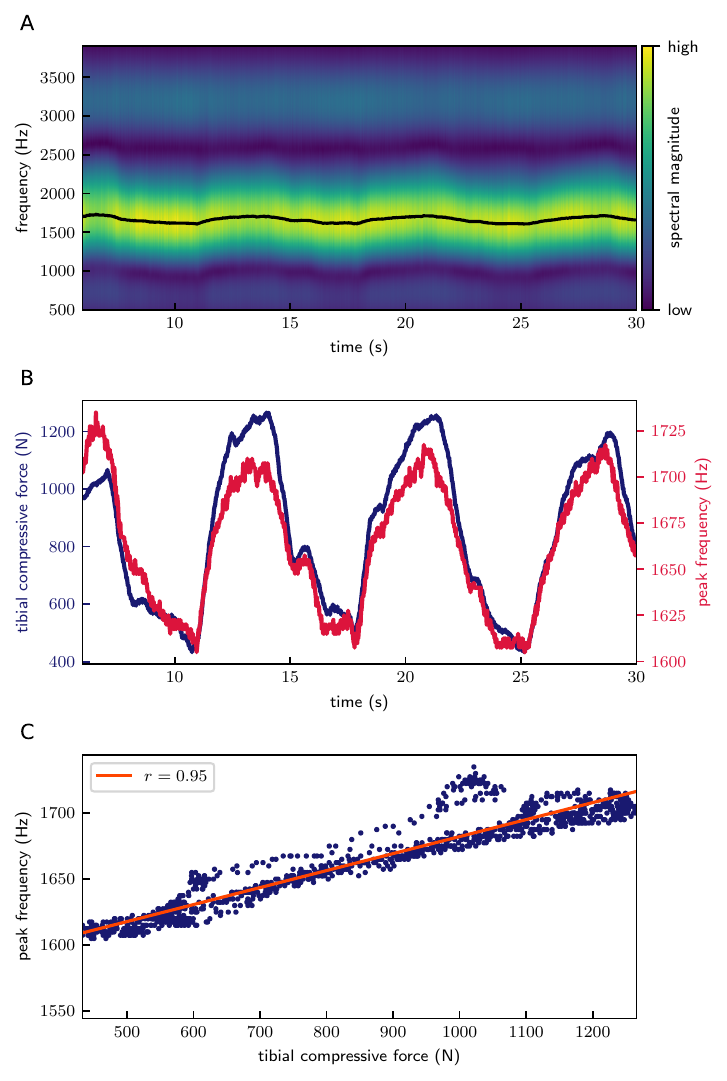}
    \caption*{Participant 9.}
    \end{subfigure}
    \caption{Data from swaying trials for all participants}
\end{figure}

\begin{figure}[h!]
  \centering
  % First Row
  \begin{subfigure}[b]{0.3\textwidth}
    \includegraphics[width=\textwidth]{figures/walkplots/results-walk-S01.pdf}
    \caption*{Participant 1}
  \end{subfigure}
  \hfill
  \begin{subfigure}[b]{0.3\textwidth}
    \includegraphics[width=\textwidth]{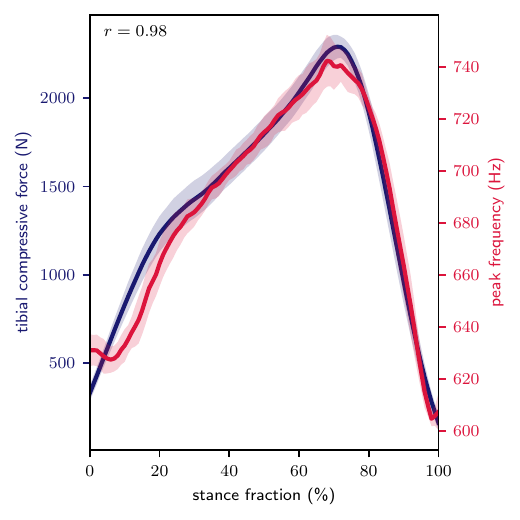}
    \caption*{Participant 2}
  \end{subfigure}
  \hfill
  \begin{subfigure}[b]{0.3\textwidth}
    \includegraphics[width=\textwidth]{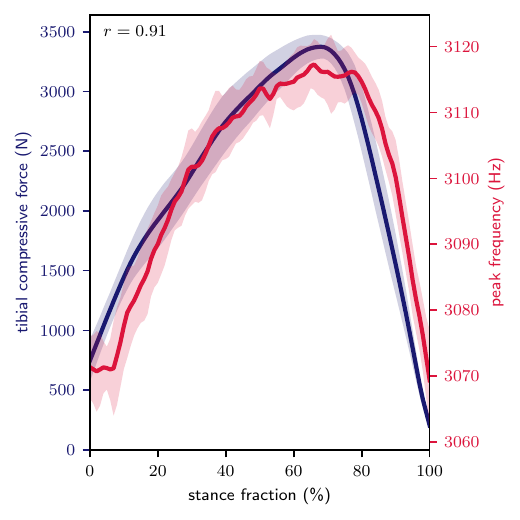}
    \caption*{Participant 3}
  \end{subfigure}

  \vspace{1cm} % Vertical space between rows

  \begin{subfigure}[b]{0.3\textwidth}
    \includegraphics[width=\textwidth]{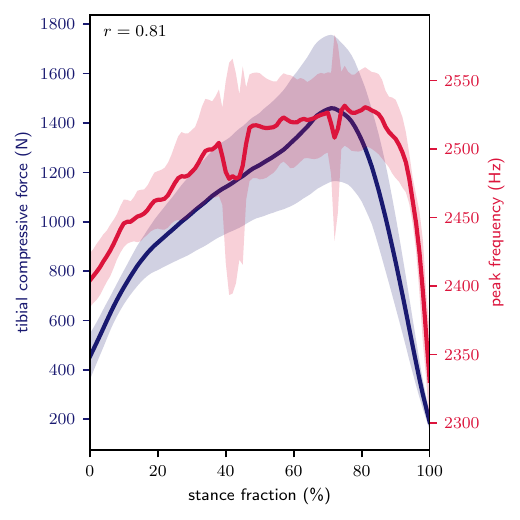}
    \caption*{Participant 4}
  \end{subfigure}
  \hfill
  \begin{subfigure}[b]{0.3\textwidth}
    \includegraphics[width=\textwidth]{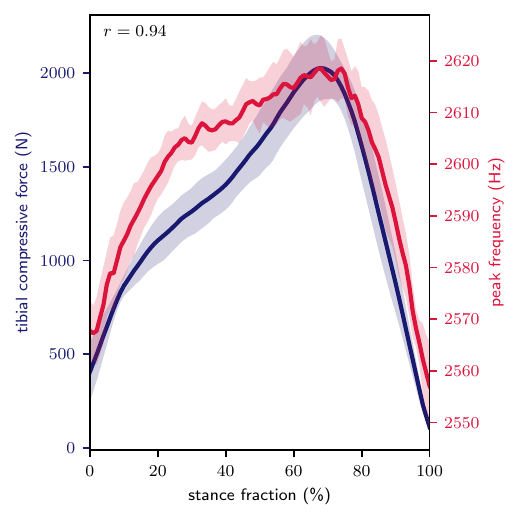}
    \caption*{Participant 5}
  \end{subfigure}
  \hfill
  \begin{subfigure}[b]{0.3\textwidth}
    \includegraphics[width=\textwidth]{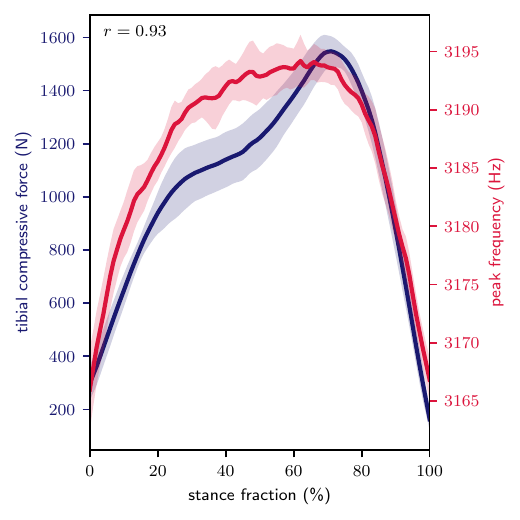}
    \caption*{Participant 6}
  \end{subfigure}

  \vspace{1cm} % Vertical space between rows

  \begin{subfigure}[b]{0.3\textwidth}
    \includegraphics[width=\textwidth]{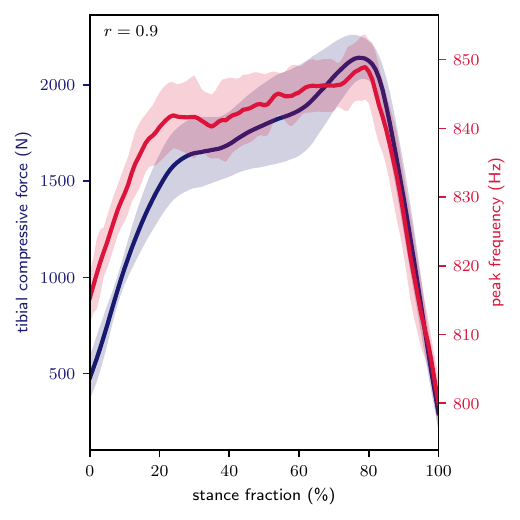}
    \caption*{Participant 7}
  \end{subfigure}
  \hfill
  \begin{subfigure}[b]{0.3\textwidth}
    \includegraphics[width=\textwidth]{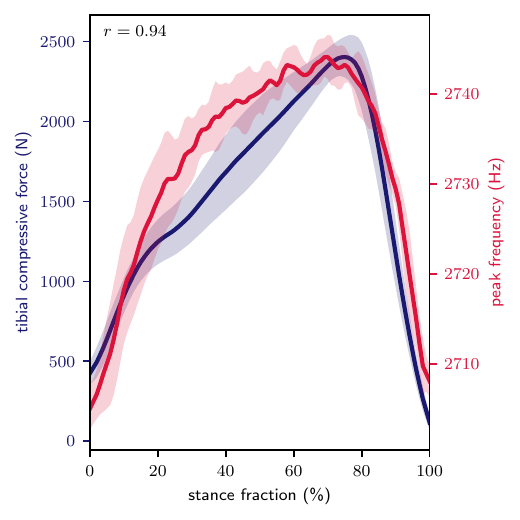}
    \caption*{Participant 8}
  \end{subfigure}
  \hfill
  \begin{subfigure}[b]{0.3\textwidth}
    \includegraphics[width=\textwidth]{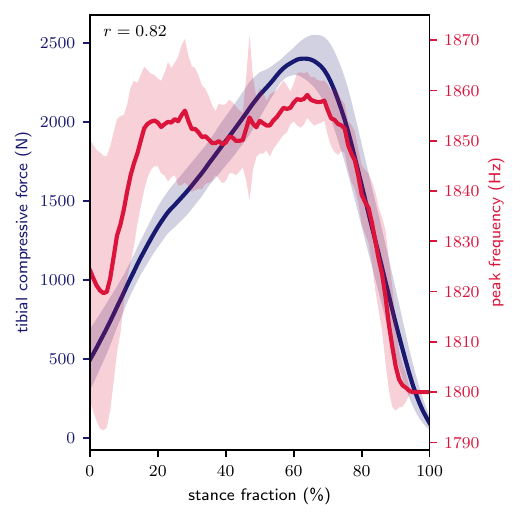}
    \caption*{Participant 9}
  \end{subfigure}

  \caption{Data from walking trials for all participants.} 
  \label{fig:walking-data-supplement}
\end{figure}

% \end{document}